# ROYAL MAUSOLEUMS
# OF THE WESTERN-HAN AND OF THE SONG CHINESE DYNASTIES: A CONTEXTUAL APPROACH

Giulio Magli
School of Architecture, Urban Planning and Construction Engineering,
Politecnico di Milano, Italy
Giulio.Magli@polimi.it

*The mausoleums of the emperors and of some members of the royal family of the Western Han Chinese dynasty – popularly known as "Chinese pyramids" - are a spectacular ensemble of tombs covered by a huge earth mounds, spread in the outskirts of modern Xian. Their inspiring model is the world-famous tomb of the first emperor Qin, who reigned immediately before the Han, and in turn they were of inspiration for the much later mausoleums of the Song dynasty. Using satellite data we investigate here on cognitive aspects of the project of these two groups of monuments, with particular attention to the problem of their orientation and of their placement in the landscape; in particular, the presence of two distinct patterns of orientation, both connected with the polar region of the sky, arises. The analysis includes the cultural relationships of astronomy with orientation and topography, as well as a – negative – test of the possible influence of the Feng Shui tradition in the Western Han period. The special, much relevant case of the funerary landscape of emperor Wen of Han is also discussed.*

## 1. Introduction

A fundamental breakthrough in Chinese history is the reign of Qin, who succeeded in unifying the country in 221 BC, becoming the first emperor. His name is worldwide famous due to the astonishing archaeological discovery of the Terracotta Army guarding his - yet unexcavated – tomb. Qin's tomb, located in Lintong not far from modern Xian, lies beneath a huge tumulus (burial mound) of compacted earth which is an unmistakable landmark of the funerary landscape of the king.
The Qin reign rapidly became unstable after his death, and was unified again in 202 BC under Liu Bang of Han (usually refereed to as Gaozu), who became thus the first emperor of the dynasty we call Western Han. The capital city was established at Changan, present day Xian.
The Western Han dynasty (206 BC – 24 AD) was a period of particular stability and wealth of the state, during which important political, economical and scientific developments occurred. The western Han rulers (with the notable exception of Wen, which will be discussed separately) followed the custom established by Qin of being buried in tombs located under huge mounds of compacted earth. The tombs – excluding the famous pits of emperor Jing's burial, where a enormous quantity of miniature terracotta figurines has been found – are unexcavated. Their identification (commonly accepted by Archaeology) is mostly based on the stelae which were erected in front of them in the second half of the 18[th] century. Much later the Western Han period, another Chinese dynasty, the Song, will revive the tradition of constructing Mausoleums in the form of burial mounds.

The mounds are a fascinating presence in the rapidly developing landscape of modern China. Most of them are accompanied by secondary burials in the form of smaller mounds, hosting relatives or members of the royal family. Taken together, these mounds constitute a impressive ensemble of monuments, constructed in a relatively short historical period to act as perennial icons of the power and the desire of eternal life of their owners (Wu 2010). In this sense, much more than in nonsensical pages available on the web, they are really the analogous of the pyramids of the 4[th] and the 5[th] Egyptian dynasties, constructed more than 2000 years before. Of course there is no cultural connections between the two; however – as the Historian of Religion Mircea Eliade has first shown - the mechanisms of the sacred and the way in which the religious power exploits himself are in many cases similar. The present paper thus presents an approach to the cognitive and contextual aspects of the project the Chinese royal mounds along the lines of research developed in recent years for the Egyptian funerary landscapes (see e.g. Magli 2011, 2012, 2013, Belmonte and Magli 2016).

**2. The Qin tomb and the Western Han burial mounds**

Beginning with the first emperor and thereafter with the Western Han dynasty, a fundamental transformation of ancient Chinese burial practices occurs (Goulong 2005). The tomb becomes a underground "palace", and the standard burial equipment (mostly of bronze ritual vessels, bells and weapons) is replaced by a variety of pottery personages (life-sized in the case of Qin) and by accurate jade suits. In particular, the underground tomb of Qin is described by the Chinese historian Sima Qian, writing a century after the First Emperor's death, as a microcosm endowed with vaults representing the heavenly bodies and a miniature of the Qin empire – including rivers of mercury – on the ground. Although Sima Qian is not always affordable, the general description of the tomb should be quite accurate, as well as the presence of mercury, as some probes have recently shown. The complex in itself was conceived as a symbolic replica of the imperial town, with above ground inner and outer enclosures, and a series of underground pits and galleries containing "replicas" of the living world familiar to the emperor. Those already excavated contain an incredible variety of brutal equipments, from bronze chariots to horses to stone armours, up to a replica pond with bronze cranes and, of course, the worldwide famous terracotta warriors. There is no doubt on the fact that the Qin building programme, and in particular that of his own tomb, served to political ends. The First Emperor credited himself as a "pivot" God, the one upon which everything else in the universe hinged. As a consequence, the tomb – whose construction began immediately after accession to the

throne – was positioned near a important mountain peak, Mt. Li, sacred to Taoists and dominating from the south the Wei River. The burial complex is signalled in the landscape by a huge mound of compacted earth, constructed probably with the help of a stepped core. The mound was in itself called mountain, and was thus meant as a replica of nature, over which the owner of the tomb exploited his power and control. The tomb's location is to the east of Qin's capital, in front of the eastern pass, so that all travellers form that direction had to cross the "funerary" city, which served also to defensive aims.

Propaganda aims were conveyed also trough the construction of the Qin capital in itself. The town was indeed conceived to convey the message of the quasi-divine nature of the emperor and of his legitimacy as universal ruler, keeper of ancient traditions and rituals. In the words of Sima Qian the layout of the imperial centre is described in analogy with the celestial realm. In particular, an explicit parallelism is made between the imperial palaces connected by a road crossing a covered bridge on the river Wei, and the "axis" - the meridian – connecting the north celestial pole with due south. There was, therefore, an overwhelming symbolic importance and concern for cardinal orientation on Earth, connected as it was with the "cosmic" order in the heavens. Actually, archaeological discoveries from the Chinese Bronze Age have demonstrated that such a dominant concern already existed during the Xia, Shang, and Zhu dynasties, forming the core organizing principles of early Chinese cosmological thinking (Pankenier 2011).

There is no doubt that a rigorous concept of cardinality inspired also the Qin funerary complex. The enclosures are oriented to the cardinal points, as well as the pits. Thousands of terracotta warriors, in their ordered lanes ready for the battle, face due east, perhaps to guard the emperor form enemy souls coming from that direction, and the burial mound is oriented cardinally as well (see next section).

The Qin dynasty collapsed rapidly at Qin's death, and the founder of the new dynasty, Gaozu, was fronted with the problem of legitimating his power trough the divine mandate of the Heaven. To express this concept trough architecture, he started a new building program which included the construction of the "city of the Heaven", the Han capital Changan.

Historical sources report that the circuit wall of the capital was inspired by the stylized form of two constellations, the Northern Dipper (Ursa Major), and the Southern Dipper (essentially our Sagittarius) (Pankenier 2011). This idea seems to be confirmed, at least in part, by archaeological surveys (Hotaling 1978). In any case, the explicit connection of the new dynasty's power with the cosmic order and therefore with cardinality is beyond any doubt.

Gaozu choose to continue the Qin tradition of the construction of burial mounds, thus establishing the standard for the dynasty that will follow. The tomb of the first Han emperor is on the vast flatland located on the the opposite side of the Wei river with respect to Mount Li. Most of the rulers who later followed selected a building site in the same area, so that still today the "pyramids" of the western Han dynasty rulers and of their relatives form a fascinating landscape, dotting as they are the course of the

Wei for some 40 Kms. The annals of the Han dynasty make it clear that the construction of the royal tomb was of extreme importance; it was a state project which involved a huge mass of people - moved to the "funerary town" which usually grew up in the vicinity of the tomb - and started early in the second year of reign. Each complex received a specific name ending in -ling, tomb.

According to the annals, the choice of the place for the tomb was a delicate matter in itself, and actually the distribution of the monuments in the necropolis is a bit puzzling. Indeed, they were not built in a linear succession from east to west along the river, but many "jumps" back and forth occurred (Fig. 1). A historical source (the Book of the Han, finished in 111 AD) mentions the use of the "Zhaomu" doctrine (essentially, alternating Zhao and Mu meaning left/right, east/west) as a method in positioning the tombs with respect to those of the two predecessors, but not all scholars agree on its application (Brashier 2011, Loewe 2016). Actually, the alternate distribution appears to apply at least to the tombs of Gaozu, Hui and Jing (if we omit the choice of a natural mountain made by Wen in between) and to those of Yan, Cheng and Ai (Fig. 1). In particular the latter three, whether by chance of by design, are also connected by a  straight line and are sufficiently close to be inter visible. Probably similar considerations of inter-visibility were applied also in the other cases, together with more practical considerations (in particular, the use of the existing facilities and workforce of the funerary towns of the predecessors).

The original form of the mounds (strictly pyramidal, or truncated pyramid) and the original heights are difficult to establish because the summits are too deteriorated. In any case, as occurs today, it was possible also in ancient times to ascend to the summit, since "ascending an imperial tomb" to offer sacrifices is documented in the Han annals as a yearly rite  exploited by the relatives of the deceased emperor. The bases of the imperial mounds are squared with the exception of the first two, which are rectangular. As far as the present author is aware, their dimensions have never been studied systematically. In absence of accurate reliefs, we can use satellite imagery, taking of course into account that the errors will be relevant,  especially because the corners of many of the mounds are not well delineated. Using these  data it is anyway possible to identify a certain regularity (Table 1). The first two mounds (Gaozu and Hui) are almost identical and about 136x170 mts; then occurs the break of emperor Wen (Section 5). After the break, the fourth emperor Jing establishes a square plan of about 170x170 mts. The long reign of emperor Wu leads to an enormous burial mound (248x248 mts), while all the following will adopt sides which are again between 160 and 170 mts except for the last emperor, Ping, whose sides are about 225 mts.

The Han unit measure of length, the *Bu,* is well known and equivalent to 1.386 Mts. Therefore, taking into account the modern errors, the dimensions of the first two mounds were probably of 100x120 Bu (138.6 x 166.3 mts) and the 120 Bu measure was adopted thereafter for most of the later ones, excluding Ping (probably 160x160 Bu) and Wu (probably 180x180 Bu).

## 3. The orientations of the Western Han burial mounds

The orientations of all the monuments under exam here have been determined (Table 1) using satellite imagery extracted from Google Earth and Bing and imported in AutoCad. In all cases the measures have been repeated on several images of the same place taken from the historical archives of the programs. Due to the high quality of the images and the low projection error associated with them (Potere 2008) the intrinsic error expected from this kind of measures is quite low. However, errors do arise mostly because of the difficulty in individuating in a precise way the sides of the mounds. Overall, we can estimate the data presented to have an uncertainty of ±1°. The first monument of interest is the huge burial mound of Qin which, as mentioned, is oriented cardinally: it has a slightly trapezoidal form, with azimuths 182° for the longitudinal side and 90° for the transversal side. For all the later monuments I will concentrate here on the azimuths of the longitudinal ("north-south") sides, and on the imperial mounds only, since the orientations of the associated mounds (which belong to members of the royal family) are always very close to that of the corresponding principal mound (complete data are reported in Table 1); interestingly, these satellite mounds do never lie on the same parallel of the main ones, so that the orientation procedure must have been repeated for each one of them.
The orientations are as follows:

1) The Gaozu mound is oriented at 167°. A close orientation is shared by his successor Hui, whose mound is identical also in shape and dimensions.
2) The third emperor Wen operates a unique break in the burial tradition, since his "mound" is a natural hill; we shall discuss this case in Section 5.
3) The fourth emperor Jing returns to the burial mound tradition, but this time he restores also the cardinal orientation of the Qin mound.
4) With the two successors Wu and Zhao we see again a macroscopic deviation from true north, although the difference is less than that of Gaozu: the azimuths are 171 and 172 respectively,
5) We return again to cardinal orientation with Xuan (180°) and Yuan (179°).
6) A final oscillation occurs with Cheng (171°) with the successors Ai and Ping who both return to strict cardinality (180°).

We have therefore two (and only two) distinct families of orientation, holding for all the imperial mounds of the Western Han dynasty:

- Family (1): precise orientation to the cardinal points, with errors not exceeding ±1°.
- Family (2): rough orientation to the cardinal points, with errors with respect to the geographic north of several degrees. The errors are always to the west of north and exhibit a tendency to decrease in time from a maximum of 14° to a minimum of 8°.

These facts cry out for an explanation, since it is absolutely clear that two different methods of orientation were used; indeed if a single method leading errors up to 15° was used, we would have data also in the interval of errors between 1° and 8°.

The mounds of family (1) were clearly oriented determining the cardinal directions. The methods used by the Chinese astronomers to find the east-west and the north-south lines are well known (Pankenier 2009).

First of all, a solar method was in use, based on the measure of the shadow of a post (gnomon) on a graduated circle on the ground. A stellar method was also used, based on the Asterism *Dìng*, the "square" of the constellation we call Pegasus. The two stars which, when connected, compose the "Eastern Wall" of such a square, namely Alpheratz and Algenib, culminate simultaneously to the south. Therefore, by projecting the "wall" perpendicularly to the ground, the surveyors could determine the geographical south and therefore the meridian. This method had also a very clear ritual content, since the divine emperor was in principle meant to be located in the north, looking therefore south along his reign.

It is possible that the two orientation methods were used *together*, the solar one for east-west orientation and the stellar one for south-north orientations, as testified by some ancient sources (Pankenier 2009) (if the two methods were used together on occasion of the Qin mound, this may explain its slight trapezoidal form).

There is no doubt whatsoever that the maximal error in determining the meridian which the builders may have committed with any of the above methods does not exceed a couple of degrees, and therefore the orientations of family (2) *cannot* be explained in this way. In the present section I will propose a solution which seems, at least to me, in optimal agreement with the way of conceiving power and the state religion during the Western Han dynasty. The reader should, however, be advised that another explanation also exists (Charvátová et al. 2011), as will be discussed in next section.

The idea is that the maximal western elongation of a circumpolar star with respect to the pole – and therefore, actually, the distance in degrees of the same star to the pole - was used. First of all in fact, we should recall that in the centuries of the Western Han dynasty no "pole star" was available, since, due to precession, the north celestial pole was not close to any naked-eye visible star. The nearest star to the pole was not our Polaris, but another star of Ursa Minor, Kochab. The dark polar region, and the closest asterism Ursa Minor, were of paramount importance for the Chinese. Its function of "pivot" of the sky was equated to the centrality of the imperial power on Earth, and the stars belonging to this region were considered as a sort of celestial emperor's palace, the Ziwei or Purple Enclosure. Each star was apparently equated with a element of the royal family inhabiting this heavenly palace. In particular and in spite of the fact that, as mentioned, Polaris was not the nearest star to the celestial pole, there is no doubt that this star was considered as extremely important (Sun and Kistemaker 1997). The Chinese called it the great emperor (or great God) of the heaven, a celestial sky god endorsing the emperor's power on Earth (it is also possible that this term was used to individuate the whole dark area of the Pole near Polaris, a thing which in any case does not change its relevance).

If we follow Polaris in its slow, apparent approach to the pole due to precession we can see that its maximal western elongation of course decreases. This elongation appears to be in rough agreement with the gradual shift in the orientations of the mound, being ~13° in 200 BC, 12° 20' in 100 BC, and 11° 40' in 1 AD. Notice that orienting to the maximal elongation (west, or east) of a star is not an easy task (so that errors of the order of one degree or more can be expected), since it implies determining the height of the pole with precision. Orienting to the maximal elongation, however, was the unique way to proceed if they wanted to "point" the mounds towards Polaris, since of course the star was circumpolar so no orientation to its rising or setting was possible.

**4. The possible relationship with the Feng Shui tradition.**

The Feng Shui is the Chinese tradition of "geomancy", that is, divination trough geographic and morphological features of the terrain. In a nutshell, the idea was that special locations exhibiting selected features were favourable from the point of view of encapsulating and enhancing "chi" (a "positive energy", allegedly flowing on the Earth) and were therefore to be electively chosen for buildings, especially tombs. Of course, it is perhaps worth stressing that the whole thing has no scientific basis whatsoever: Feng Shui is considered here because of its relevant historical importance only.
Two main "schools" of Feng Shui appear to have existed, a "Form" and a "Compass" school. The Form school is based on the observation of the morphology of the landscape ( mountains and rivers) and their relationship to winds. The main idea is that a "correct place" must first of all have a 'Principle Mountain', the tallest mountain in the surroundings, to the north, and a conspicuous river at the southern boundary. Starting from these two main characteristics, a series of intricate further refinements and analyses can be applied to establish the supposed level of suitableness of a site.
The compass school is instead strictly related to the Chinese invention of the lodestone compass, which is mentioned - *as a orientation tool* - in a text written before the end of the fourth century BC (Needham 1959). Apparently then, the discovery of the spontaneous alignment of a freely floating stick of lodestone was at a certain point interpreted by Feng Shui specialists as a mean to establish the direction of flow of the alleged "vital energy". As a consequence, the started to be used to identify – with respect to magnetic north - a set of directions each one having its "geomantic" character.
In spite of ubiquitous claims of a very ancient origin of both kinds of Feng Shui geomancy, to the best of the author's knowledge it is not possible to identify written sources about it before the 3 century AD. In spite of this, the tradition must have been existing - and the compass was certainly existing – in the period of interest here. Thus, it certainly makes sense to investigate if it was applied in the project of the imperial tombs during the Western Han period. It is worth mentioning that a similar hypotheses has been put forward also for the orientation of Maya temples and cities; however, in such a case proofs of knowledge of the compass in Pre-Columbian times are very weak and there is no cultural evidence available, while - on the contrary - evidences for astronomical/calendrical orientations in the Maya world are overwhelming (see e.g. González-García and Šprajc 2016).

Let us first analyse landscape ("Form") aspects. The test is definitively and immediately *un*successful: the Qin tomb seems to having been thought of exactly at the opposite of the main Feng Shui rule, since the principal mountain (Mount Li) is to the south, and the river Wei is to the north. As far as the burial mounds of the western Han dynasty are concerned, they are located in plain flatland; the river Wei flows to the south but no kind of mountain curtain occurs nearby in any other direction, so one should eventually interpret the mounds themselves as the "principal mountains".

What about the possible use of the Compass Feng Shui or anyway, the use of a compass for orienting the mounds? The problem here is that modelling the Earth's magnetic field in the ancient past – a thing which is fundamental if we want to know the direction indicated by a compass at the time of construction of each mound – is very difficult. In spite of this, a model curve of the magnetic declination as a function of time is available, although affected by a relevant error of ±5°. It has been shown that, when the orientations of the mounds are reported as points on the same Cartesian plane, those belonging to family (2) – namely those skewed many degrees to the north west – appear to follow the general slope of the curve (Charvátová et al. 2011). As a consequence, the use of the magnetic compass is a feasible, alternate explanation for the orientations of family (2). However, at least in the present author view, the cultural and contextual arguments discussed in the previous section, united with the fact that there is no evidence for "geomantic" use of the compass during the Western Han dynasty, rather point for the astronomical solution.

Regarding family (1), it should be noticed that, since the bandwidth of error in determining magnetic declination is so large, also the mounds of this family fall into the realm of possible compass orientation (see again Charvátová et al. 2011). However, it must be taken into account that the error in estimating the magnetic declination affects the positioning of the curve within the bandwidth, and it is not a physical error which can be associated to a measure point by point. Therefore, including also the cardinal orientations makes the behaviour of the orientations inconsistent (Fig. 3). To better explain the point, consider for instance the royal mounds of Yuan, Cheng and Ai. As already mentioned, the three are located on a straight line and are very close: the length of the line is less than 7 Kms. However, Yuan's mound is cardinally oriented; Cheng's mound – whose project was laid on the terrain 16 years later – is skewed 10° to the west, and finally Ai's, designed on the ground 26 years later, is again cardinally oriented. Clearly, if all the three have been aligned magnetically, this would correspond to a almost random behaviour of the magnetic declination at the place.

**5. The funerary landscape of emperor Wen**

The reign of the third Han emperor, Wen, is singular under many respects. In particular, he is known for the reform of the empire, and for his "frugality" and attention to the people. Wen's reign is peculiar also because he is the unique Western Han ruler who did not build a burial mound. The project of his tomb – called Baling - thus differs considerably from the imperial tombs that had preceded it. For his tomb, Wen selected a mountain, and the funerary chambers were hollowed out of the rock under the mountain.

The tomb was not constructed in the same area of his two predecessors, north of the Wei River (where in any case there are no suitable mountains) but instead to the south of the river and of the capital, selecting a natural hill which is today signalled on the ground by stone stelae of subsequent dynasties (the interior is unexcavated). Wen's mountain tomb is quite important in the history of Chinese tomb architecture since several elite burials later imitated its style, including those of many local kings of the Western Han period.

In the past, several theories have been put forward to explain the design and location of emperor Wen's tomb. One such theories calls in question the fact that emperor Wen could not have been rightfully buried in the region north of the Wei because he was the brother of the former emperor and not his heir. Besides being doubtful, this idea does not explain the choice of the natural mountain instead of the mound.

The fact that the tomb is moundless in turn has been sometimes explained as a sign of moderation and frugality, to avoid the huge expenses and efforts required for a mound-tomb, and by a desire of security. Again, these explanations appear doubtful, as the tomb is reported to having been looted in the past, and to contain a very rich burial equipment. A huge funerary town was founded on the occasion, and excavating the tomb must have been quite a huge project on his own. A more sound explanation is that the unconventional choice was a way of communicating identity and authority of the ruler, acting as a unifying pivot for the members of the royal clan (Miller 2015). Symbolism, therefore, was the main reason for the unusual, but spectacular, choices made by Wen. The "possession" of the mountain was a powerful symbol of kingship: as the first emperor Qin some 70 years before, also Wen conducted sacrifices in the mountains near the capital, and made the local kings responsible for conducting sacrifices in the mountains in their territory. As a side bonus, the tomb's town played a role for defensive purposes, being in control of the Southern Pass.

Within the framework of Wen choices for his funerary landscape, an interesting fact can be observed here. Two of the relatives of Wen, the wife Empress Dowager Bo, and the daughter Empress Dou, have their own burial mounds. These are huge, almost identical rectangular structures - almost the size of a royal mound, around 154x135 mts (probably 110x100 Bu) – located in the plain to the south-west of Baling. They are the unique mounds of the Western Han dynasty whose orientation is not even roughly cardinal, being 113.5° (Dou's longest side is rotated 90° with respect to Bo). The solution is very probably that the tomb of the Bo was orientated towards the Baling peak, which is (barely) visible looking from the summit along the projection of the longest sides of the mound, some 3.6 km far away; the tomb of Dou was perhaps a replica of the other were the alignment was not functional.

Interestingly enough, also another emperor choose to be buried south of the river Wei, Xuan. As far as the present author is aware, no explanation has ever been proposed for the unconventional location of Xuan burial mound either. It may be noticed, however, that Xuan personality has been passed on as a hard working and brilliant emperor, who grew up as a commoner, and was keen to the needs of people. If this is true, perhaps an explicit reference to Wen's burial - and therefore, to the same values and politics – was meant in selecting the building site of Xuan's tomb. In this respect it can be noticed that – whether by chance or by design - the diagonal of Xuan's mound passes quite neatly on the apex of Baling, which was (barely) visible at the horizon, some 11.5 Kms away.

## 6. From Western Han to the Song dynasty

The tradition of building huge burial mounds was to be revived five hundreds years later by emperor Wen of Sui (581–604 AD), the founder of the Sui dynasty. Wen has been a very important historical personage, under many aspects comparable to the first emperor, and was able to reunify China after almost 3 century of chaos. Apparently, he choose to make an explicit reference to the "golden age" of the western Han dynasty with his tomb. In fact Wen's tomb is not by chance located along the river Wei, some 50 Kms further west with respect to the western Han necropolis. It is signalled by a huge mound which is cardinally oriented and whose dimensions are precisely those "standard" of the Western Han dynasty.

The Sui were rapidly replaced by one of the most enduring Chinese dynasties, the Tang. The founder, Gaozu of Tang (618–626 AD), abdicated in favour of his son Taizong. The historical sources report what happened at Gaozu death. Taizong explicitly ordered that the construction of Gaozu mausoleum had to be inspired by Changling, the mausoleum of Gaozu of Han, the first Emperor of the Han. The tomb was built on the ridge of the Xian flatland, and is in fact characterized by a huge, cardinally oriented mound. The tomb is not far from the mountains which were instead selected for Taizong's own tomb, Zhaoling. In fact, regarding his own tomb, the emperor explicitly stated that it had to be inspired by Baling, the mountain tomb of Emperor Wen of the Han, and selected for it the peak of Mount Jiuzong. The king's mausoleum consisted (besides the underground tomb, which is yet unexcavated) in a series of architectural projects and, in particular, in a monumental processional way, the so-called *spirit path,* flanked by huge statues and forming with the mountain a spectacular ensable. Taizong inaugurated in this way the tradition of constructing spectacular mountain-tomb mausoleums which all the later Tang emperors will follow, starting from the emperor's son, Gaozong, who built Quianling, an astounding masterpiece of the interplay between nature and man-made architecture.

After the Tang, another period of great splendour for China was the (Northern) Song dynasty (960-1127 AD), with capital in Bianjing (Kaifeng). The seven imperial mausoleums of the Song Dynasty are located in western Gongyi, Henan, some 120 Kms to the east of the capital (a further mausoleum was constructed east of the first one for the father of the first emperor, who was posthumously proclaimed emperor Xuanzu). The topography of the area (Fig. 4) clealry shows that they were disposed according to a rule. This rule is not, however, the Zhaomu we have seen previously, but rather, they are arranged in couples, with the later element being slightly to the north-west of the previous. Interestingly, the unique exception is the tomb of emperor Zhenzong, which looks isolated However, to the north-west of it (marked E in figure 4) there is the unique satellite Song mausoleum of a certain relevance, endowed – in particular – with a complete spirit path. It is the tomb of empress Liu of Song. She served as regent of China during illness of the husband (1020-1022 AD), and later during the minority of the son, Emperor Renzong.

The placement of the monuments in the general landscape is (as for Qin mausoleum) in "inverse" Feng-shui rules since the river is to the north of the flat plain and (not particularly prominent) mountains are located to the south. As far as orientation is concerned, the burial mounds are notably smaller than those of the Han and in poorer conditions, but all of them are reached from the south by spectacular, straight spirit paths lined by statues representing ministers and generals but also lions, elephants and other subjects. In Table 2 we thus report the orientations of the Spirit Paths for the seven imperial Mausoleums (the Xuanzu is very difficult to measure; the orientation of Empress Liu is close to that of Zhenzong at ~185°)

Once again, we see a clear intention to cardinality, as seven out of seven orientations fall between 180° and 185°. Also in this case, a rough agreement with the reconstructed local magnetic declination is observable (Charvátová et al. 2011). During the Song dynasty, relevant advances in understanding the scientific concept of magnetic declination are documented, however, and on the other end the north celestial pole was slowly completing its approach to Polaris, the distance being reduced to ~6° during the 150 years spanned by the seven Song emperors. It is thus very conceivable that also these monuments were oriented to the maximal (in this case east) elongation of Polaris.

## 7. Discussion and conclusions

The burial mounds of the western Han emperors are a wonderful ensemble of monuments, which stand still today as imposing icons, pointing to the divine power of their owners. They form an impressive sacred landscape in which placement and orientation helped to convey political messages. In particular, we have investigated about the role of the relationship between the divine nature of the emperor – the pivot of the kingdom - and the north celestial pole/the stars of Ursa Minor in the orientations of the mounds. These orientations are very clearly divided in two families: one is certainly connected with precise – albeit ritual – method of finding true south, and therefore true north, trough the stars. The second might be connected with more "esoteric" traditions but appears to be more easily and naturally explicable if an orientation to Polaris is supposed and followed trough the precessional "motion" of this star in the Western Han period.

The tradition of building mausoleums based on burial mounds is later followed by the emperors of the Song dynasty, and also in this case a clear attention to the spatial relationships between the monuments and with the northern region of the sky emerges. The cultural landscape of the royal mounds in China survived the trial of the centuries, and it has to be hoped that it may survive as much as possible also to the needs of such a rapidly developing country.

## Table 1
## Mausoleums of Qin and of the Western Han dynasty

| Map | Emperor/ date of reign | Associated burials | Long. Az. | Tras. Az. | Name of tomb and dimensions (ns/ew) | Notes |
|---|---|---|---|---|---|---|
| Q | Qin | | 182 | 90 | 368x362 | |
| 1 | Gaozu 202-195 BC | | 167 | 77 | Changling 135x168 | |
| | | Emp. Lü | 167 | 77 | | SE of main |
| 2 | Hui 195–188 BC | | 166 | 75 | Anling 137x170 | |
| | | Emp. Zhang | 167 | 77 | | NW |
| | | Marq. Zhang | 166 | 76 | | NE |
| | | Princ. Lu | 165 | 76 | | NE |
| W | Wen 180–157 BC | | | | Baling | Natural hill |
| | | Emp. Dou | 114 | 22 | | Towards Baling |
| | | Emp. Bo | 114 | 22 | | |
| 3 | Jing 157–141 BC | | 179 | 89 | Yanling 170x170 | |
| | | Emp. Wang | 179 | 89 | | NE |
| 4 | Wu 141–87 BC | | 172 | 81 | Maoling 248x248 | |
| | | Emp. Li | 172 | 81 | | NW |
| 5 | Zhao 87-74 BC. | | 171 | 82 | Pingling 160x160 | |
| | | Emp. Shangguan | 171 | 82 | | NW |
| 6 | Xuan 74–49 BC | | 180 | 90 | Duling 170x170 | Opposite river of Wei |
| | | Emp. Wang | 180 | 90 | | SE |
| | | Emp. Xu | 180 | 90 | | SE |
| 7 | Yuan 49–33 BC | | 179 | 90 | Weiling 170x170 | |
| | | Emp. Wang | 179 | 90 | | NW |
| 8 | Cheng 33–7 BC | | 170 | 80 | Yangling 160x160 | |
| | | Emp. Xu | 171 | 81 | | NW, with 7 other small mounds |
| | | Cons. Ban | 170 | 80 | | NE |
| 9 | Ai 7–1 BC | | 180 | 90 | Yiling 160x160 | |
| | | Emp. Fu | 177 | 87 | | NE |
| 10 | Ping 9 BC-AD 6 | | 180 | 90 | Kanling 225x225 | |

## Table 2
## Mausoleums of the Song dynasty

| Map | Emperor/date of reign | Long. az. (spirit path) | Name of tomb | |
|---|---|---|---|---|
| 1 | Xuanzu d. 956 | | Yongan | |
| 2 | Taizu 960–976 | 183 | Yongchang | |
| 3 | Taizong 976–997 | 185 | Yongxi | |
| 4 | Zhenzong 997–1022 | 184 | Yongding | |
| 5 | Renzong 1022–1063 | 184 | Yongzhao | |
| 6 | Yingzong 1063–1067 | 184 | Yonghou | |
| 7 | Shenzong 1067–1085 | 183 | Yongyu | Final section bends to 178° |
| 8 | Zhezong 1085–1100 | 180 | Yongtai | |

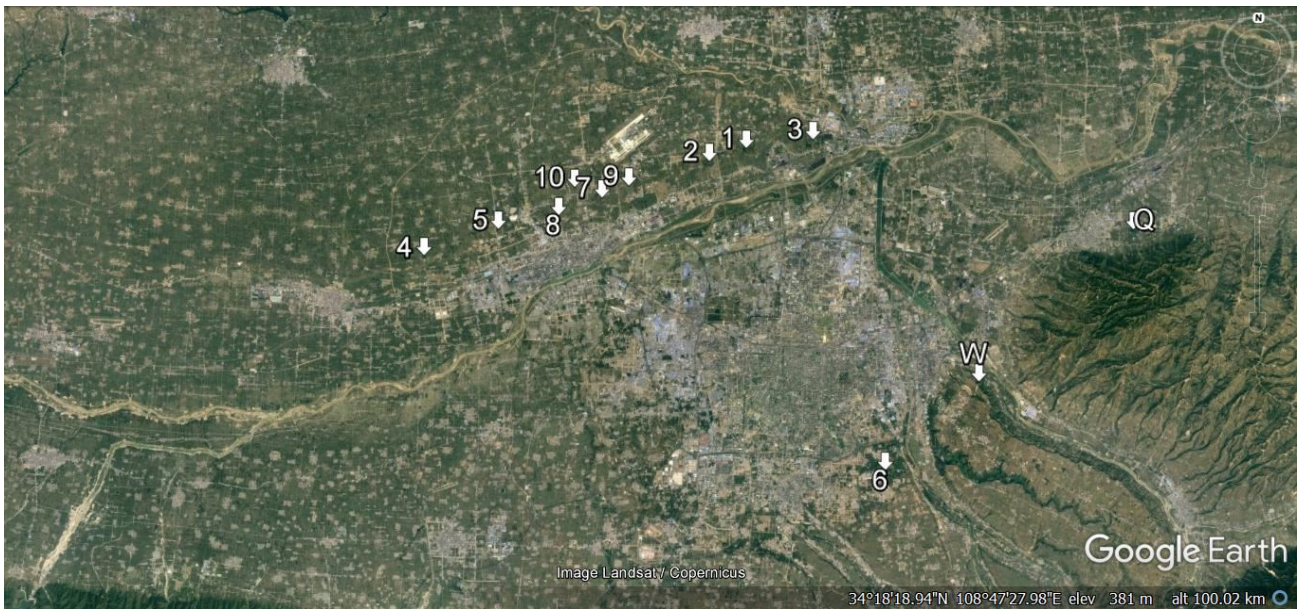

Fig. 1
The burial mounds of the Western-Han dynasty. 1 Gauzu Chanling, 2 Hui Anling, 3 Jing Yanling, 4 Wu Maoling, 5 Zhao Pingling, 6 Xuan Duling, 7 Yuan Weiling, 8 Cheng Yangling, 9 Ai Yiling, 10 Ping Khanling. Letters Q and W denote the positions of Qin and of Wen (Baling) tombs (Image courtesy Google Earth, editing by the author)

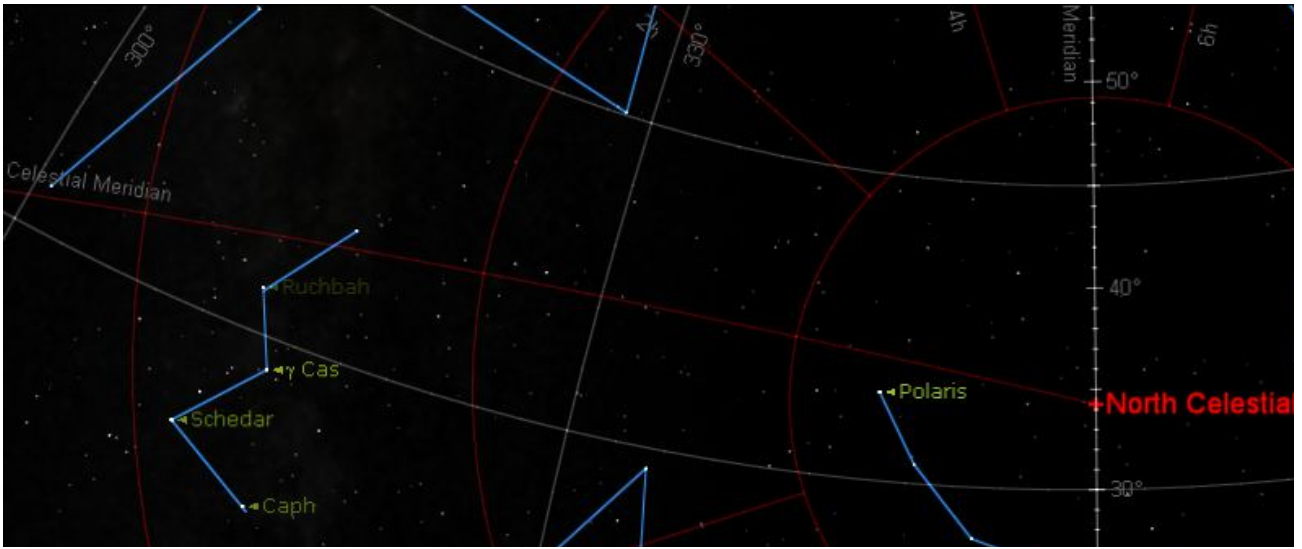

Fig. 2
The region of the North celestial pole as seen from Xian in 206 BC, with Polaris at maximal western elongation.

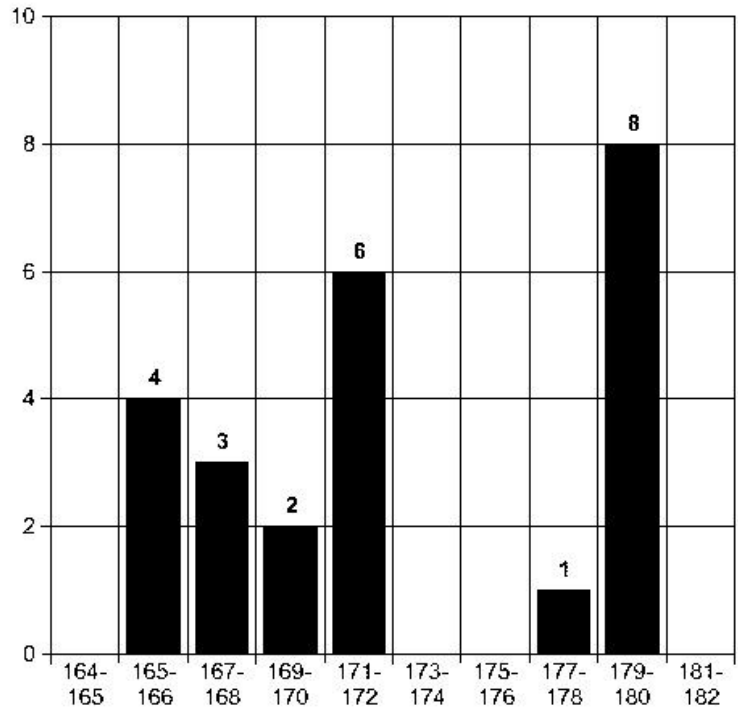

Fig. 3
Orientation histogram of the 24 measurable burial mounds of the Western Han dynasty with a 2° x-axis interval. No mounds fall out of the range 165°-180°. Further, a clear gap occurs between 172° and the cardinal orientation.

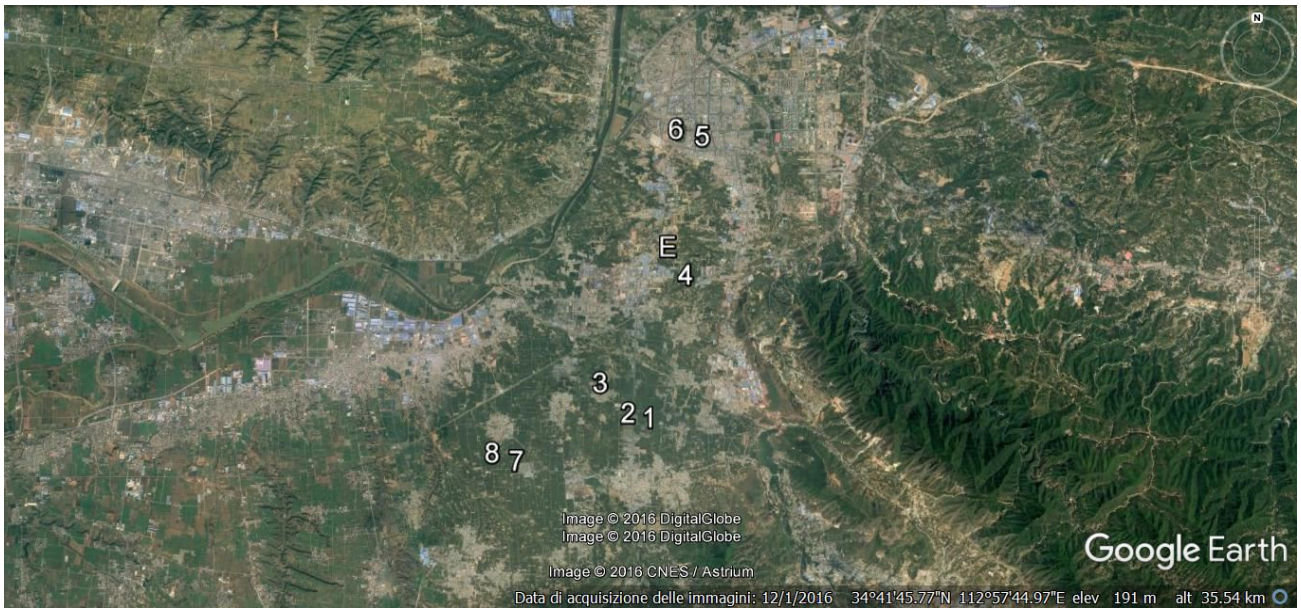

Fig. 4
Mausoleums of the Song dynasty. 1) Xuanzu Yongan, 2) Taizu Yongchang, 3) Taizong Yongxi, 4) Zhenzong Yongding, 5) Renzong Yongzhao, 6) Yingzong Yonghou, 7) Shenzong Yongyu, 8) Zhezong Yongtai, E) Empress Lu of Song (Image courtesy Google Earth, editing by the author)